\documentclass[twocolumn,showpacs,preprintnumbers,amsmath,amssymb]{revtex4}

\usepackage{graphicx}
\usepackage{dcolumn}
\usepackage{bm}

\newcommand\beq{\begin{eqnarray}}
\newcommand\eeq{\end{eqnarray}}

\def\kvec{\mbox{\boldmath $k$}_\perp}
\def\Kvec{\mbox{\boldmath $K$}_\perp}
\def\k3vec{\mbox{\boldmath $k$}}

\def\lvec{\mbox{\boldmath $l$}_\perp}

\def\vepvec{\mbox{\boldmath $\vep$}_\perp}

\def\Delvec{\mbox{\boldmath $\Delta$}_\perp}
\def\0vec{\mbox{\boldmath $0$}_\perp}

\def\vep{\varepsilon}

\def\slash#1{\rlap/{#1}}

\def\qslash{\slash{\mkern-1mu q}}

\def\vepslash{\slash{\mkern-1mu \varepsilon}}
\def\Delslash{\slash{\mkern-1mu \Delta}}

\def\Qbar{\overline{Q}}

\input{epsf.sty}
\input{epsfig.sty}

\begin{document}


\title{
Investigation of the color-dipole structure in diffractive $t$-slopes 
of charmonia photo- and electroproductions
}

\author{A. Hayashigaki$^{(1)}$
and K. Suzuki$^{(2)}$
}
\address{
$^{(1)}$ 
Institut f\"ur 
Theoretische Physik, Universit\"at Regensburg,
D-93053 Regensburg, Germany}
\email{arata.hayashigaki@physik.uni-regensburg.de}
\address{$^{(2)}$ Division of Liberal Arts, 
Numazu College of Technology, Shizuoka 410-8501, Japan}
\email{ksuzuki@la.numazu-ct.ac.jp}

\date{\today}

\begin{abstract}
The diffractive $t$-slope, $B_V$, of elastic charmonia ($V=$$J/\psi$, $\psi'$) 
photo- and leptoproductions off a nucleon is studied at low $|t|$ ($<$ 1 GeV$^2$)
in the leading logarithmic approximation of perturbative QCD, 
with a special emphasis on the space-time evolution of the $c\bar{c}$-dipole.
We obey a framework based on QCD factorization,
which describes a certain Fermi motion effect
due to the $c(\bar{c})$-quarks in the proper manner \cite{HST2,HST3} 
and includes appropriately kinematical corrections 
of the momentum transfer $\Delvec$ in the $t$-channel.
Assuming the universal two-gluon form factor of the nucleon,  
we show that the difference of $t$-slopes for $J / \psi $ and $\psi '$ 
is dominated by the contribution from the dipole-charmonium transition process.
The calculated difference is found to be $B_{J/\psi}-B_{\psi'} \sim 0.53$ GeV$^{-2}$
for the photoproduction, in agreement with HERA data.
We also calculate the $t$-dependence of the total cross sections for several
center-of-mass energies $W$.
A good agreement of the results with the available data demands that
the mass scale appearing in the gluon form factor should significantly 
decrease with increasing $W$.

\end{abstract}


\maketitle


\section{Introduction}
Recent measurements of high-energy diffractive 
photo- and electroproductions of charmonia off a nucleon, 
$\gamma^{(*)}+N(P) \rightarrow V+N(P')$ ($V = J/\psi, \psi'$), 
are  unique sources of information about both structures 
of the target nucleon and the production process of charmonia.
The $t$-distribution of differential cross sections, 
where $t$ is a square of momentum transfer to the vector meson  
defined as $\Delta=P-P'$, {\it i.e.}, 
$t=\Delta^2\simeq -\Delvec^2$, 
provides us with information concerning the spatial structure
in the plane perpendicular to the photon-nucleon reaction axis.
In this case, one can study the impact parameter distribution 
of the gluon in the nucleon, as well as the space-time evolution 
of the dipole-like $c\bar{c}$ state, 
which is created initially by the photon's fluctuation 
and hadronizes to charmonia
after a direct scattering with the gluons inside the nucleon, in the transverse space.
The former is never accessible 
in the well-known inclusive deep-inelastic scattering, 
from which one can extract only parton distributions of the nucleon 
with respect to the longitudinal momentum fraction.  
%
%
In the limit of large photon virtuality $Q^2$, 
a remarkable prediction by QCD factorization tells us
that the $t$-dependence of a production cross section 
is solely determined by a universal two-gluon form factor of the 
nucleon, 
independent of the type of vector mesons produced \cite{BFGMS}.

On the other hand, the effect of the dipole dynamics 
on the $t$-distribution of cross sections is considered less 
important than the one due to the nucleon structure, 
because the $c\bar{c}$ state is highly squeezed at large $Q^2$ \cite{BFGMS}
and, naively speaking, can be regarded as an almost point-like state 
during the whole process. 
In the diffractive charmonia photoproductions observed 
at the HERA experiment \cite{H100,ZEUS02,H102}, 
however, only a hard scale is provided by the charm quark mass $m_c$, 
{\it e.g.}, $Q^2\sim m_c^2$. 
In this case, the initial $c\bar{c}$ state is dominated 
by transverse polarizations of the quasi-real photon 
and thus should not be so squeezed.
Therefore, the evolution of the $c\bar{c}$ state in the transverse space 
most likely gives a non-negligible 
contribution to the $t$-distribution of the production amplitudes.     
It is a significant work
to estimate how the $t$-distribution of cross sections 
is sensitive to the structure of the $c\bar{c}$ state as functions 
of $Q^2$ and $W$.

For this purpose, it is natural that we formulate the cross section 
at finite $t$ along the familiar frameworks based on QCD factorization 
\cite{FKS,RRML,HST2}, which give a good description for the diffractive $J/\psi$ 
photo- and electroproductions.
The perturbative QCD analyses (pQCD) for the $J/\psi$ photoproduction 
at $t=0$ suggest a strong suppression of the total cross sections 
due to the Fermi motion of the $c(\bar{c})$-quarks, 
especially relative transverse motion between $c$ and $\bar{c}$.
In particular, it is more pronounced for a radially excited $\psi'$$(2S)$ production, 
because the radial size of the $c\bar{c}$-dipole in the final hadronic state is about 
twice as large as that of $J/\psi$ and then the transverse motion 
of the $c(\bar{c})$-quarks could be more active owing to the existence of 
a node in the $\psi'$ wave function \cite{HST2}.
Thus, detailed information concerning the internal structure 
({\it i.e.}, motion or radial size) 
of the $c\bar{c}$ state in the dipole-charmonium transition process
are considered to be inherent in the cross sections 
of the diffractive $J/\psi$ and $\psi'$ productions. 
One possible method to extract such information suitably from these processes 
would be to calculate the $t$-distributions of the cross sections.
This fact, further, motivates us to find out information associated 
with only the dipole structure of the $c\bar{c}$ state 
by making use of both the $J/\psi$ and $\psi'$ production cross sections.
This would be carried out in the following way:
experiments at HERA have observed the $t$-distribution
intensively for the $J/\psi$ photoproductions \cite{H100,ZEUS02} 
and recently, also for the $\psi'$ case \cite{H102}.
Each $t$-slope, $B_V$, extracted as a result of an exponential fit
to the data, $\mbox{exp}(B_V t)$, includes both information on the nucleon 
and the $c\bar{c}$ state, as mentioned above.
We assume that the $t$-slope is of a simple form described by a sum
of a universal nucleon form factor and the contribution from 
the dipole state at large $W$ (see Sec.~II~A).
Therefore, a {\em difference} of $t$-slopes between $J/\psi$ and $\psi'$ is 
free from the contribution of the universal nucleon structure, and 
is determined only by the finite size effect of the $c\bar{c}$-dipole.  
This difference is in fact calculated as the convolution of the dipole 
scattering amplitudes with the wave functions of $J /\psi$ and $\psi '$.  
%
%

%
Motivated by these interests, in this paper, 
we calculate a differential cross section as a function 
of $|t|$ ($<$ 1 GeV$^{-2}$)
in diffractive (elastic) charmonia ($J/\psi$ and $\psi'$) 
photo- or electroproductions off a nucleon. 
We employ familiar QCD factorization formulae with the helicity representation
in the Brodsky-Lepage approach for hard exclusive processes \cite{LB}.
It is calculated under the leading logarithmic approximation (LLA) of pQCD, 
which is reasonable for the interaction 
of a small transverse-size $c\bar{c}$-dipole.
A reliable prediction for the $t$-slopes requires a sophisticated pQCD 
analysis for the space-time evolution of the $c\bar{c}$-dipole, 
particularly giving a precise description of the Fermi motion 
between $c$ and $\bar{c}$. 
On this point, our approach in the dipole picture developed in \cite{HST2,HST3}
could be more appropriate than those in Ref.~\cite{FKS,NNPZZ}, 
because our formulation deals with $O(v^2)$ corrections due 
to the Fermi motion effect in a proper manner, 
where $v$ is an average velocity of the (anti-)charm quarks 
in the charmonium rest frame.
The previous approaches \cite{FKS,NNPZZ} are not satisfactory 
in view of reliable estimation of the Fermi motion effects, which 
should take into account 
a proper projector onto the $S$-wave wave function \cite{HST2,HST3}, 
the contribution of off-shellness in the spinor matrix elements 
of $c\bar{c}$ state and other $O(v^2)$ corrections \cite{HST3}.
The approach of \cite{FKS} is 
especially inadequate
for the description of $\psi'$,  
because their light-cone wave functions (LCWF) of charmonia are constructed
in an oversimplified fashion, such that the dependence 
of the ratio of $\psi'$ to $J/\psi$ cross sections on $Q^2$ at $t=0$
fairly underestimates the HERA data at small $Q^2$ \cite{HP,SHIAH,HST2}. 
%

%
Another development of our model is to include a kinematical correction
from finite $\Delvec$ due to a recoil of the target nucleon, 
which should explicitly appear in the arguments of 
nonperturbative charmonium wave functions 
$\Psi _V (\alpha,\kvec,\Delvec)$, 
with $\alpha$ and $\kvec$ being the longitudinal momentum 
fraction and the transverse momentum of the $c$-quark, respectively.      
Previous works \cite{FKS,NNPZZ} simply take the charmonium wave function 
$\Psi _V (\alpha,\kvec)$ in the limit of $\Delvec^2=0$
and miss kinematical corrections arising from the finite $\Delvec$.  
Since the $\kvec$-dependence of the wave function is dominated by the region, 
where $| \kvec |$ is less than the inverse of typical hadronic size, $\sim 0.5$ GeV, 
%
the contribution of $|\Delvec|$ is not small compared to $|\kvec|$, 
even if we restrict the region of $|t|=\Delvec^2$ 
to less than 1 GeV$^{2}$ in our analysis.
%
We explicitly deal with such a correction under the LLA.

With these improvements, 
we derive the $t$-slopes of both $J/\psi$ and $\psi'$ productions 
and investigate the $Q^2$ and $W$ dependences respectively.
We find that the contribution of the $c\bar{c}$ state to the $t$-slope
is not negligible and the sizes reach about $5 \sim 10$ $\%$ of the whole, 
which become largest at $Q^2=0$ for the $J/\psi$ production, 
and at $Q^2\sim 10$ GeV$^2$ for the $\psi'$ one.
The difference between $t$-slopes of $J /\psi $ and $\psi '$ is then 
calculated to study the dynamics of the $c\bar{c}$-dipole.  
We emphasize that this quantity 
provides us with an opportunity for direct comparison of the 
pQCD calculation of the dipole contribution 
with the experimental data.   
The result is actually consistent with the data 
from HERA at $Q^2=0$ \cite{H100,H102,ZEUS02}. 
%
%
%

We also calculate the $t$-dependent differential cross section of $J/\psi$
by combining the dipole part and the universal gluon form factor of the nucleon.  
We fix a value of the mass scale appearing in the dipole form of the gluon 
form factor so as to reproduce the observed $t$-dependence \cite{ZEUS02}.  
Precise determination of the mass scale should be possible, 
because the present work already gives a 
realistic description of the dipole contribution to the cross section.
In order to obtain a good fit to the data of $W$-dependence, 
it turns out that the mass scale of the gluon form factor 
should become significantly smaller with increasing $W$. 
The paper is organized as follows:
in Sec.~II, we give a detailed description of our formulae in the LLA,
focusing on the $\Delvec$ dependence.
Sec.~III contains our numerical results of the $t$-distribution
coming from the $c\bar{c}$-dipole in the $J/\psi$ and $\psi'$ 
productions. 
Here, the $Q^2$ and $W$ dependences of each $t$-distribution
and their difference are shown and
compared with HERA data at $Q^2=0$.
Also, the $W$-dependence of the mass scale of the form factor
is discussed through fits to data. 
%
%
In Sec.~IV, a summary and discussion are presented.

\section{Formulation}
\subsection{Description of $t$-slope in QCD factorization}
In the pQCD analysis,
the corresponding diffractive amplitude 
at high $W$ can be interpreted as a sequence of several steps separated 
in time, as demonstrated in \cite{BFGMS,FKS}. 
It schematically has a factorized form
\begin{eqnarray}
{\cal M}&=& \Psi_V^* \otimes 
{\cal A}_{c\bar{c}gg}
\otimes\ \tilde{\Gamma}
\otimes \Psi_\gamma.
\label{eqn:I-1}
\end{eqnarray}
Here $\Psi_\gamma$ is the LCWF of a photon
describing the $c\bar{c}$ fluctuation from the photon, ${\cal A}_{c\bar{c}gg}$ is
the hard scattering amplitude of the $c\bar{c}$-pair
off the nucleon via two-gluon exchange in the $t$-channel, 
$\tilde{\Gamma}$ represents the $t$-dependent gluon distribution in the nucleon 
and $\Psi_V$ is the charmonium LCWF 
including the soft hadronization process of the $c\bar{c}$ state.
The calculation for the transition $\gamma^{(*)}\rightarrow V$ 
is based on a technique developed in \cite{HST2, HST3}, 
which accounts for moderate sub-leading effects to $O(v^2)$ 
due to the Fermi motion of the $c\bar{c}$ state, especially important for $\psi'$.
On the other hand, the process-independent gluon distribution 
$\tilde{\Gamma}(x_1,x_2,\lvec^2,t)$
in the nucleon is assumed to be a product 
of the off-forward (skewed) unintegrated gluon distribution $f(x_1,x_2,\lvec^2)$
with a universal two-gluon form factor $\Gamma(t)$,
{\it i.e.}, $\tilde{\Gamma} = f(x_1,x_2,\lvec^2) \times \Gamma(t)$ \cite{RSS,MRT}.
Here, $x_1$ and $x_2$ denote the longitudinal momentum fractions of outgoing 
and incoming gluons respectively, and $\lvec$ the transverse 
momentum of the outgoing gluon (see Fig.~\ref{fig:fig1}).  
The form factor $\Gamma(t)$ has the same expression as a usual electromagnetic one 
of the nucleon, 
so called "dipole form" $\Gamma(t)=1/(1-t/M^2)^2$ with the mass scale $M$ \cite{FS}.
Here, it should be noted that the cross sections of these processes are 
written as the quadratic 
form of the nucleon form factor multiplied by the charmonia LCWF.
Thus, the diffractive charmonium productions could be 
sensitive to those nonperturbative objects.
The experimental differential cross section $d\sigma/dt \propto |{\cal M}|^2$ is 
usually parametrized as ${\rm exp}(B_V t)$ with the diffractive slope $B_V$,
as already mentioned.
The $t$-slope is extracted by the logarithmic derivative 
of the cross section over $t$.
Owing to the factorization ansatz, the calculated $t$-slope can be completely separated 
into two contributions 
of the nucleon part, $B_N(t)$, related to the nucleon form factor $\Gamma(t)$, 
and the dipole part, $B^{dip}_V(t)$, associated with the 
transverse evolution of the $c\bar{c}$-dipole 
in the transition process $\gamma^{(*)}\rightarrow V$:
\beq
B_V(t) = B_N(t) + B^{dip}_V(t),
\label{eqn:I-3}
\eeq
where $B_N(t)=4/(|t|+M^2)$.

As a result, 
the difference of $t$-slopes between $J/\psi$ and $\psi'$ productions
is independent of the universal nucleon form factor and thus equals the difference
of the dipole parts, 
\beq
B_{J/\psi}(t)-B_{\psi'}(t)
=B^{dip}_{J/\psi}(t)-B^{dip}_{\psi'}(t) .
\label{eqn:I-3a}
\eeq
This expression enables us to investigate the structure 
of the $c \bar c$ dipole in detail 
through a direct comparison with the data.  
%
The naive 
geometrical interpretation may suggest that 
the difference $B^{dip}_{1S}-B^{dip}_{2S}$ is negative,
because the smaller the size of color-singlet 
objects (or $q\bar{q}$ pair), 
which is approximately the inverse of mass of the vector meson,
the smaller the $t$-slopes \cite{RSS}. 
As emphasized for charmonia productions in \cite{NNPZZ}, 
however, when one considers the production 
of the radially excited state such as $\psi'(2S)$, 
such a naive expectation is broken because of the node effect 
in the $2S$ wave function. 
Actually, recent HERA data at the photoproductions seem to
support $B_{J/\psi} > B_{\psi'}$ \cite{H102}.

\subsection{Differential cross section in the LLA of pQCD}
\begin{figure}[h]
\vspace*{0cm}
\begin{center}
\hspace*{0cm}
\psfig{file=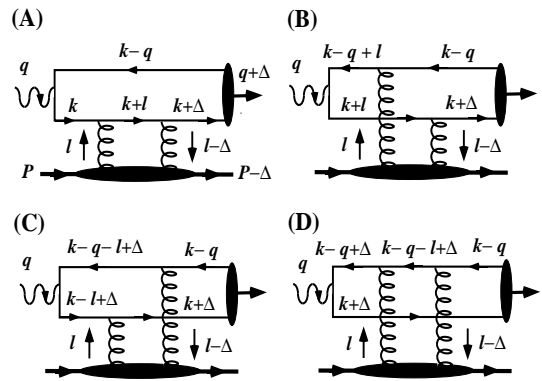,width=7.5cm,height=5cm}
\end{center}
\vspace{0cm}
\caption{Four Feynman diagrams for 
$\gamma^{(*)}+N \rightarrow V(=J/\psi,\psi')+N$ with momentum labeling.}
\label{fig:fig1}
\end{figure}

In the LLA, we formulate the diffractive 
photo- or electroproduction of charmonia off the nucleon, 
$\gamma^{(*)} (q)+ N (P) \to V (q+\Delta) + N (P- \Delta)$ $(V=J/\psi,\psi')$, 
where $P$ and $q$ denote the four momenta of the photon and the incoming nucleon, respectively, and $\Delta$ the momentum transfer in the $t$-channel, as illustrated in Fig.~\ref{fig:fig1}.   
The total center-of-mass energy of the $\gamma^{(*)}$-$N$ system, 
$W = \sqrt{(P + q)^2}$, is then assumed to be much larger than the photon's virtuality $Q^2$ and the heavy-quark mass $m_c$.
Following the usual fashion, we perform 
the standard Sudakov decomposition of all momenta 
labeled in Fig.~\ref{fig:fig1}, by introducing two null vectors $q'$ and $p'$, 
{\it i.e.}, $q = q'-(Q^2/s)p'$,
$P = (M_N^2/s)q'+p'$,
$k = \alpha q'+\beta p' + k_\perp$,
$l = \kappa q'+x_1 p' + l_\perp$ and 
$\Delta = (t/s)q'+\delta x\ p'+\Delta_\perp$. 
Here, $s = 2q'\cdot p' \simeq W^2-M_N^2+Q^2$ 
with the masses of the nucleon and charmonium, $M_N$ 
and $M_\psi$, respectively, and $\Delvec^2=-t$. 
We assume that $q^{\prime +}$ and $p^{\prime -}$ are much larger than 
$q^{\prime -}$ and $p^{\prime +}$ in our reference frame.
The coefficients $\beta$ and $\kappa$ are of the order $1/s$. 
$x_1$ has the relation
$x_1=(M_{c\bar{c}}^2+Q^2)/s+O(\lvec)$ with the squared mass of 
the intermediate $c\bar{c}$ state, 
$M_{c\bar{c}}^2= (\kvec^2+m_c^2)/[\alpha(1-\alpha)]$.
The second term of $x_1$, $O(\lvec)$, corresponds to the contribution 
beyond the LLA, and thus we neglect this term hereafter. 
The longitudinal momentum fraction of the incoming gluon, $x_2=x_1-\delta x$, is 
given as $x_2 \simeq (M_{c\bar{c}}^2-M_\psi^2+t)/s$.
Since $x_2 \ll x_1$ in the small region of $|t|<1$ GeV$^2$, 
this means the off-forward (or skewed) kinematics for the two-gluon exchange.
%
%

The differential cross section with the small momentum transfer $t$
to the nucleon has the form
\beq
\lefteqn{\frac{d\sigma(\gamma^{(*)}N \rightarrow VN)}{dt}}
\nonumber\\
&=&\frac{1}{16\pi W^4}\sum_{pol.=L,T}\left|{\rm Im} {\cal M}^{pol.}(t)
+i\,{\rm Re}{\cal M}^{pol.}(t)\right|^2,
\nonumber\\
\label{eqn:II-1}
\eeq
where the suffix '$pol.$' denotes the polarizations of the incoming photon.
The imaginary part of the amplitude ${\cal M}^{pol.}$ is of the form
\beq
\lefteqn{{\rm Im}{\cal M}^{pol.}(t)} 
\nonumber\\
&=& \frac{\pi s}{\sqrt{2N_c}}
\int_0^1 d\alpha \int\frac{d\kvec}{16\pi^3}\int d\lvec
\frac{\alpha_s(\lvec^2)}{\lvec^2(\lvec-\Delvec)^2}
\nonumber\\
&\times& \tilde{\Gamma}(x_1,x_2,\lvec^2,t)
\, I^{pol.(\gamma\rightarrow V)}(\alpha,\kvec,\lvec,\Delvec)
\label{eqn:II-2}
\eeq
with the number of color $N_c=3$ and $\alpha_s=g^2/(4\pi)$, 
where $I^{pol.(\gamma\rightarrow V)}$
represents the transition $\gamma^{(*)}\rightarrow V$ such as
\beq
\lefteqn{I^{pol.(\gamma\rightarrow V)}(\alpha,\kvec,\lvec,\Delvec)}
\nonumber\\
&=& \sum_{\lambda\lambda'} \left[
\Psi^{(A)}_{\gamma}(\alpha,\kvec)-\Psi^{(B)}_{\gamma}(\alpha,\kvec,\lvec)
\right.
\nonumber\\
&&\left.-\Psi^{(C)}_{\gamma}(\alpha,\kvec,\lvec,\Delvec)
+\Psi^{(D)}_{\gamma}(\alpha,\kvec,\Delvec)
\right]_{\lambda'\lambda}
\nonumber\\
&& \times\ \Psi_{V\lambda\lambda'}^{*}(\alpha,\kvec,\Delvec)
\label{eqn:II-3}
\eeq
with the helicities of charm and anti-charm quarks, $\lambda'$ and $\lambda$, respectively.
The function $\tilde{\Gamma}(x_1,x_2,\lvec^2,t)$ describing 
the lower part of Fig.~\ref{fig:fig1}, as mentioned in Sec.~II~A, has the form 
$\tilde{\Gamma}= f(x_1,x_2,\lvec^2) \Gamma(t)$ with the two-gluon form factor,
\beq
\Gamma(t)=\frac{1}{\left(1-t/m_{2g}^2\right)^2},
\label{eqn:II-a}
\eeq
where $m_{2g}$ denotes a mass scale relevant to this process \cite{FS}. 
The skewed unintegrated gluon distribution $f(x_1,x_2,\lvec^2)$ 
is related with
\beq
f(x_1,x_2,\lvec^2) = 
\frac{\partial x_2 G(x_1,x_2,\lvec^2)}{\partial {\rm ln}\ \lvec^2},
\label{eqn:II-4}
\eeq
using the skewed gluon distribution $G(x_1,x_2,\lvec^2)$.
The effect of skewedness can be then incorporated into a constant 
enhancement factor $R$ at small $x_1$ like the HERA experiment \cite{MR}. 
Therefore, we rewrite the 
function $f(x_1,x_2,\lvec^2)$ in terms of usual diagonal gluon distribution as
\beq
f(x_1,x_2,\lvec^2) = \frac{R\ \partial x_1 G(x_1,\lvec^2)}{\partial {\rm ln}\ \lvec^2} .
\label{eqn:II-5}
\eeq
Here, $R= 1.1 \sim 1.2$ for the $J/\psi$ photoproduction \cite{MR} 
and this factor gives an enhancement by $20 \sim 40$ $\%$ for the cross section. 
Thus, we express the imaginary part of amplitude in terms of the diagonal 
gluon distribution, where hereafter we denote $x_1$ as $x$ for simplicity.
Then, the real part in Eq.~(\ref{eqn:II-1}) is related to the imaginary part as 
${\rm Re}{\cal M} = -(\pi/2)(\partial\,{\rm Im}{\cal M})/(\partial\, {\rm ln}x)$ 
in the perturbative analysis \cite{BFGMS}.

\subsection{Light-cone wave function of photon}
First, let us construct the LCWF of the incoming photon $\Psi_\gamma$ 
in Eq.~(\ref{eqn:II-3}), 
based on the LC perturbation theory \cite{LB,BFGMS}. 
It is defined as the following spinor matrix element 
\beq
\lefteqn{\Psi_{\gamma\lambda'\lambda}^{pol.}(\alpha,\kvec,\lvec,\Delvec)}
\nonumber\\
&=&\frac{\overline{u}_{\lambda'}(p_i)}{\sqrt{\alpha}}\vepslash_\gamma^{pol.}
\frac{v_{\lambda}(q_i)}{\sqrt{1-\alpha}}
\frac{-e e_c \alpha(1-\alpha)}{\kvec^{(i)2}+m_c^2+\alpha(1-\alpha)Q^2},
\nonumber\\
\label{eqn:II-6}
\eeq
where $\varepsilon_\gamma^{pol.}$ denotes
$\varepsilon_\gamma^0=(q'+x p')/Q$ 
for the longitudinal polarization, and 
$\varepsilon_\gamma^\perp=(0,0, \vepvec^\gamma)$ with
$\vepvec^\gamma=\frac{1}{\sqrt{2}}(1,\gamma i)$ and $\gamma=\pm 1$
for the transverse polarization. 
Here, $\varepsilon_\gamma^{pol.}\cdot q
=\varepsilon_\gamma^0\cdot\varepsilon_\gamma^\perp=0$.
$e_c$ is the charge of charm quark $2/3$ in the units of $e$. 
$p_i$ and $q_i$ are the momenta of charm and anti-charm quarks 
with the helicities $\lambda'$ and $\lambda$ respectively, for each diagram $(i)$ 
illustrated in Fig.~\ref{fig:fig1}. 
We define $\kvec^{(i)}$ as $\kvec^{(A)}=\kvec$, 
$\kvec^{(B)}=\kvec+\lvec$, $\kvec^{(C)}=\kvec-\lvec+\Delvec$ 
and $\kvec^{(D)}=\kvec+\Delvec$ for the corresponding diagram.
After calculation of the spinor matrix elements,
the results of the longitudinal polarization are given by
\beq
\Psi_{\gamma\lambda'\lambda}^{long.} &=& 0 
\label{eqn:II-7}
\eeq
for $\lambda'=\lambda$, and
\beq
\Psi_{\gamma\lambda'\lambda}^{long.}
&=& \frac{e e_c}{Q}\left[1-\frac{2\alpha(1-\alpha)Q^2}
{\kvec^{(i)2}+m_c^2+\alpha(1-\alpha)Q^2}\right]
\label{eqn:II-8}
\eeq
for $\lambda'=-\lambda$.
Here, the factor 1 in the first term of Eq.(\ref{eqn:II-8}) disappears, 
when one takes a sum of four diagrams in calculation of the amplitude, 
and thus one neglects the factor.
The results of the transverse polarization are 
\beq
\Psi_{\gamma\lambda'\lambda}^{trans.(\gamma)} 
&=& 
-\frac{\sqrt{2} e e_c \lambda \delta_{\gamma,\lambda} m_c }
{\kvec^{(i)2}+m_c^2+\alpha(1-\alpha)Q^2}
\label{eqn:II-9}
\eeq
for $\lambda'=\lambda$, and
\beq
\Psi_{\gamma\lambda'\lambda}^{trans.(\gamma)}=
-\frac{2 e e_c (\alpha-\delta_{\lambda,\gamma})
\vepvec^{\gamma}\cdot\kvec^{(i)}}{\kvec^{(i)2}+m_c^2+\alpha(1-\alpha)Q^2}
\label{eqn:II-10}
\eeq
for $\lambda'=-\lambda$. 

\subsection{Light-cone wave function of charmonium}
Next, we describe the LCWF of the outgoing charmonium produced through 
the soft hadronization process of the $c\bar{c}$ state, $\Psi_V$. 
It is defined in analogy with that of a photon. However, one should pay 
attention to new contribution from non-zero momentum of $\Delvec$, 
different from the case of the photon. 
Namely, the center of mass motion of the charmonium 
deviates by $\Delvec$ to a direction perpendicular 
to the incoming photon, due to a recoil of the target nucleon.
Therefore, we introduce new null vectors $\bar{q'}$ and $\bar{p'}$ 
to redefine $(q+\Delta)^+$ 
as the longitudinal direction of $\Psi_V$.
Here, $\bar{q'}$ and $\bar{p'}$ 
are expressed as $\bar{q'}=q'-(t/s)p'+\Delta_\perp$ 
and $\bar{p'}=-(t/s)q'+p'-\Delta_\perp$ 
up to $O(1/s)$ in terms of $q'$ and $p'$. 
The momenta of two quarks in the $c\bar{c}$ state, $k+\Delta$ and $k-q$, are written as 
\beq
k+\Delta &=& \alpha \bar{q'} + \left[\frac{\alpha t}{s}+\beta
+\frac{Q^2-t+M_\psi^2}{s}\right] \bar{p'}
\nonumber\\
&&+k_\perp+(1-\alpha)\Delta_\perp,
\nonumber\\
k-q &=& (\alpha-1) \bar{q'} + \left[\frac{\alpha t}{s}+\beta
+\frac{Q^2-t}{s}\right] \bar{p'}
\nonumber\\
&&+k_\perp+(1-\alpha)\Delta_\perp .
\label{eqn:II-11}
\eeq
Using these relations, we find that the relative momentum 
between $c$ and $\bar{c}$ is $k_\perp+(1-\alpha)\Delta_\perp$.
The polarization vectors of the charmonium 
are also rewritten in terms of $q'$ and $p'$ as 
$\varepsilon_\psi^0=[q'-(M_\psi^2+t)p'/s+\Delta_\perp]/M_\psi$ 
for the longitudinal polarization, and 
$\varepsilon_\psi^\perp=\varepsilon_\gamma^\perp+2
(\Delta_\perp\cdot \varepsilon_\gamma^\perp)(q'-p')/s$
for the transverse one.
Here, $\varepsilon_\psi^0\cdot (q+\Delta)
= \varepsilon_\psi^\perp\cdot (q+\Delta)
= \varepsilon_\psi^0\cdot\varepsilon_\psi^\perp=0$.
Using those polarization vectors,
the LCWF of the charmonium $\Psi_V$ has the form
\beq
\Psi_{V\lambda\lambda'}^{pol.*}(\alpha,\kvec,\Delvec)
&=&\frac{\overline{v}_\lambda(q-k)}{\sqrt{1-\alpha}}\vepslash_\psi^{pol.*}
{\cal R} \frac{u_{\lambda'}(k+\Delta)}{\sqrt{\alpha}}
\nonumber\\
&\times&\frac{\phi_\psi^*(\alpha,\kvec+(1-\alpha)\Delvec)}{M_\psi} ,
\label{eqn:II-12}
\eeq
where ${\cal R}$ represents the projector introduced in \cite{HST2,HST3}, 
which keeps the $c\bar{c}$ state to be low-lying $1S$ state, given as
${\cal R}=\left[1+(\qslash+\Delslash)/M_\psi\right]/2$. 
Following \cite{HST3}, we further take into account the off-shellness 
of the spinor matrix element,
$\overline{v_{\lambda}}\vepslash_\psi^{pol.*}{\cal R} u_{\lambda'}$,
as the proper treatment.
This spinor matrix element representing the effective 
``$c\bar{c}\rightarrow V$ vertex'' may be off the energy shell 
and thus the total energy carried by the two spinors 
$\bar{v}_{\lambda}$ and $u_{\lambda'}$ is generally different 
from the energy of the vector meson. 
This contribution should give $O(v^{2})$ corrections in the final result
(In more detail, see \cite{HST3}). 
For the longitudinal polarization, the results lead to
\beq
\Psi_{V\lambda\lambda'}^{long.*}
&=&-\sqrt{2}\lambda (1-2\alpha)
\vepvec^{(\lambda)}\cdot(\kvec+(1-\alpha)\Delvec)
\nonumber\\
&\times&
\frac{\phi_\psi^*(\alpha,\kvec+(1-\alpha)\Delvec)}{2\alpha(1-\alpha)M_\psi}
\label{eqn:II-13}
\eeq
for $\lambda'=\lambda$, and
\beq
\Psi_{V\lambda\lambda'}^{long.*}
&=&-\frac{1}{2}\left[1+\frac{(\kvec+(1-\alpha)\Delvec)^2+m_c(M_\psi+m_c)}
{\alpha(1-\alpha)M_\psi^2}\right]
\nonumber\\
&\times& \phi_\psi^*(\alpha,\kvec+(1-\alpha)\Delvec)
\label{eqn:II-14}
\eeq
for $\lambda'=-\lambda$, 
where $\vepvec^{(\lambda)}=\frac{1}{\sqrt{2}}(1,\lambda i)$.
For the transverse polarization, similarly,
\beq
\Psi_{V\lambda\lambda'}^{trans.(\gamma')*}
&=& \sqrt{2}\lambda \left[\delta_{\gamma',\lambda}
\left\{(M_\psi+m_c)m_c+\alpha(1-\alpha)M_\psi^2\right\}
\right.
\nonumber\\
&&\left.+2\delta_{\gamma',-\lambda}
\left\{\vepvec^{\gamma' *}\cdot(\kvec+(1-\alpha)\Delvec)\right\}^2
\right]
\nonumber\\
&\times&
\frac{\phi_\psi^*(\alpha,\kvec+(1-\alpha)\Delvec)}{2\alpha(1-\alpha)M_\psi^2}
\label{eqn:II-15}
\eeq
for $\lambda'=\lambda$, and
\beq
\Psi_{V\lambda\lambda'}^{trans.(\gamma')*}
&=& 2\vepvec^{\gamma' *}\cdot(\kvec+(1-\alpha)\Delvec)
\nonumber\\
&\times&
\left[(\alpha-\delta_{\gamma,\lambda})(M_\psi+2m_c)+(1-2\alpha)m_c\right]
\nonumber\\
&\times&
\frac{\phi_\psi^*(\alpha,\kvec+(1-\alpha)\Delvec)}
{2\alpha(1-\alpha)M_\psi^2}
\label{eqn:II-16}
\eeq
for $\lambda'=-\lambda$. The suffix $\gamma'$ denotes the polarization 
$\gamma'=\pm 1$ in the transverse direction.

\subsection{Detailed derivation of differential cross section}
We focus on the process of $S$-channel helicity conservation (SCHC)
between the initial photon and the final charmonium, 
because the HERA experiment demonstrates that, to good accuracy,
the hypothesis of SCHC holds 
in the kinematical range $30 < W < 200$ GeV 
and $|t|< 1$ GeV$^{-2}$ \cite{ZEUS02}. 
We obtain the following expressions for $I^{pol.(\gamma\rightarrow V)}$ 
of Eq.~(\ref{eqn:II-3}), using Eqs.~(\ref{eqn:II-7})-(\ref{eqn:II-10}) and (\ref{eqn:II-13})-(\ref{eqn:II-16}):
for the longitudinal channel of $\gamma(L)\rightarrow \psi(L)$,
\beq
I^{L\rightarrow L} &=& 2 e e_c \alpha(1-\alpha)Q\ \Phi_1
\nonumber\\
&\times&
\left[1+\frac{(\kvec+(1-\alpha)\Delvec)^2+m_c(M_\psi+m_c)}
{\alpha(1-\alpha)M_\psi^2}\right]
\nonumber\\
&\times& \phi_\psi^*(\alpha,\kvec+(1-\alpha)\Delvec)\ ;
\label{eqn:II-17}
\eeq
for the transverse channel of $\gamma(T)\rightarrow \psi(T)$,
\beq
I^{T(\gamma)\rightarrow T(\gamma)} &=&
-\frac{e e_c}{\alpha(1-\alpha)M_\psi^2}
\nonumber\\
&\times&
\left[m_c\left\{(M_\psi+m_c)m_c+\alpha(1-\alpha)M_\psi^2\right\}
\Phi_1\right.
\nonumber\\
&-&2\left.\left\{(2\alpha(1-\alpha)-1)M_\psi-m_c\right\}
\vepvec^\gamma\cdot\vec{\Phi}_2\right. 
\nonumber\\
&&\times
\left.\vepvec^{\gamma *}\cdot(\kvec+(1-\alpha)\Delvec)\right]
\nonumber\\
&&\times \phi_\psi^*(\alpha,\kvec+(1-\alpha)\Delvec)\ .
\label{eqn:II-18}
\eeq
Here, conveniently 
we defined the sum of energy denominator as
\beq
\Phi_1 &=& \frac{1}{\kvec^2+\Qbar^2}
-\frac{1}{(\kvec+\lvec)^2+\Qbar^2}
\nonumber\\
&-&\frac{1}{(\kvec+\Delvec-\lvec)^2+\Qbar^2}
+\frac{1}{(\kvec+\Delvec)^2+\Qbar^2}\ ,
\nonumber\\
\label{eqn:II-19}
\\
\vec{\Phi}_2 &=& \frac{\kvec}{\kvec^2+\Qbar^2}
-\frac{\kvec+\lvec}{(\kvec+\lvec)^2+\Qbar^2}
\nonumber\\
&-&\frac{\kvec+\Delvec-\lvec}{(\kvec+\Delvec-\lvec)^2+\Qbar^2}
+\frac{\kvec+\Delvec}{(\kvec+\Delvec)^2+\Qbar^2}\ ,
\nonumber\\
\label{eqn:II-20}
\eeq
where $\Qbar^2=m_c^2+\alpha(1-\alpha)Q^2$. 

Now we are in a position to perform analytical
integrations over $\lvec$ and $\kvec$ in the amplitude (\ref{eqn:II-2}).
First, we explain the integration over $\lvec$, 
assuming the range of $\lvec^2$ 
relevant in this process as
$\Delvec^2 \ll \lvec^2 \ll \Qbar^2+\kvec^2$.
This permits one to make an approximation $1/(\lvec-\Delvec)^2\simeq 1/\lvec^2$ 
for the gluon propagators in Eq.~(\ref{eqn:II-2}). 
Then, the $\lvec$-integration for $\Phi_1$ in the integrand of 
Eq.~(\ref{eqn:II-2})
leads to 
\beq
\lefteqn{\int_{\mu^2}^{\Qbar^2+\kvec^2} d\lvec
\frac{\alpha_s(\lvec^2)}{\lvec^2}
\frac{\partial xG(x,\lvec)}{\partial \lvec^2} \Phi_1(\alpha,\kvec,\lvec,\Delvec)}
\nonumber\\
&\simeq& \pi\alpha_s(\Qbar^2+\kvec^2) xG(x,\Qbar^2+\kvec^2)
\ F_1(\alpha,\kvec,\Delvec)
\nonumber\\
\label{eqn:II-21}
\eeq
with 
\beq
\lefteqn{F_1(\alpha,\kvec,\Delvec) =
-\frac{1}{\left(\kvec^2+\Qbar^2\right)^2}
+\frac{2\Qbar^2}{\left(\kvec^2+\Qbar^2\right)^3}}
\nonumber\\
&-&\frac{1}{\left\{(\kvec+\Delvec)^2+\Qbar^2\right\}^2}
+\frac{2\Qbar^2}{\left\{(\kvec+\Delvec)^2+\Qbar^2\right\}^3}.
\nonumber\\
\label{eqn:II-22}
\eeq
Here, we integrated from the low energy cut-off $\mu^2$($\sim \Lambda_{QCD}^2$) to 
$\Qbar^2+\kvec^2$ \cite{RRML,MR} over $\lvec^2$.
Similarly, for $\vepvec^\gamma\cdot \vec{\Phi}_2$ in the integrand of 
Eq.~(\ref{eqn:II-2}), we get 
\beq
\lefteqn{\int_{\mu^2}^{\Qbar^2+\kvec^2} d\lvec
\frac{\alpha_s(\lvec^2)}{\lvec^2}
\frac{\partial xG(x,\lvec)}{\partial \lvec^2} 
\vepvec^\gamma\cdot \vec{\Phi}_2(\alpha,\kvec,\lvec,\Delvec)}
\nonumber\\
&\simeq& \pi\alpha_s(\Qbar^2+\kvec^2) xG(x,\Qbar^2+\kvec^2)\ 
\vepvec^\gamma\cdot \vec{F}_2(\alpha,\kvec,\Delvec)
\nonumber\\
\label{eqn:II-23}
\eeq
with 
\beq
\lefteqn{\vepvec^\gamma\cdot \vec{F}_2(\alpha,\kvec,\Delvec)
=}
\nonumber\\
&&2\Qbar^2 \left[
\frac{\vepvec^\gamma\cdot\kvec}{\left(\kvec^2+\Qbar^2\right)^3}
+\frac{\vepvec^\gamma\cdot(\kvec+\Delvec)}{\left\{(\kvec+\Delvec)^2+\Qbar^2
\right\}^3}\right],
\nonumber\\
\label{eqn:II-23}
\eeq
where we neglected $O\left((\lvec^2/\Qbar^2)^2\right)$ terms in the LLA.
Next, we analytically perform  
the angular integration over $\kvec$ in Eq.~(\ref{eqn:II-2}). 
We introduce a new integral variable, 
$\Kvec=\kvec+(1-\alpha)\Delvec$, instead of $\kvec$,
to rewrite the nonperturbative wave function 
$\phi_\psi(\alpha,\kvec+(1-\alpha)\Delvec)$ 
as a function of one variable, $\Kvec$.
Here, we stress that, as mentioned in Sec.~I, 
the contribution of $\Delvec$ is not suppressed compared with that of $\kvec$ 
in the argument of wave function,
and thus we should exactly include the kinematical corrections of $\Delvec$
by such a redefinition of the argument.
When we define the angle $\theta$ as 
$\Kvec\cdot\Delvec=|\Kvec||\Delvec|\,{\rm cos}\,\theta$, 
the $\theta$-integrations of $F_1$ and $\vec{F}_2$ 
yield the following results:
\beq
\lefteqn{\int_0^{2\pi}d\theta F_1(\alpha,\Kvec,\Delvec)}
\nonumber\\
&=&-\frac{4\pi}{\left(\Kvec^2+\Qbar^2\right)^2}
\nonumber\\
&&\times
\left[1+\frac{2\{(\alpha^2+(1-\alpha)^2)\Delvec^2-\Qbar^2\}}{\Kvec^2+\Qbar^2}
\right.
\nonumber\\
&&-\frac{12(\alpha^2+(1-\alpha)^2)\Delvec^2\Qbar^2}
{\left(\Kvec^2+\Qbar^2\right)^2}
\nonumber\\
&&+\left.\frac{12(\alpha^2+(1-\alpha)^2)\Delvec^2\Qbar^4}{\left(\Kvec^2+\Qbar^2\right)^3}\right],
\label{eqn:II-24}
\eeq
\beq
\lefteqn{\int_0^{2\pi}d\theta\ \vepvec^\gamma\cdot
\vec{F}_2(\alpha,\Kvec,\Delvec)\ 
\vepvec^{\gamma *}\cdot \Kvec}
\nonumber\\
&=&\frac{4\pi\Kvec^2\Qbar^2}{\left(\Kvec^2+\Qbar^2\right)^3}
\left[1
+\frac{3(\alpha^2+(1-\alpha)^2)\Delvec^2}{\Kvec^2+\Qbar^2}\right.
\nonumber\\
&& 
-\left.\frac{6(\alpha^2+(1-\alpha)^2)\Delvec^2\Qbar^2}
{\left(\Kvec^2+\Qbar^2\right)^2}\right].
\label{eqn:II-25}
\eeq
Here, we left the $t$-dependent terms up to 
$O\left(\Delvec^2/(\Kvec^2+\Qbar^2)\right)$. 
Therefore, in the photoproduction, this implies  
that the applicable range of $t$ should be $|t| \ll m_c^2$ and
we should discuss numerical results in the small regions of $|t| < 1$ GeV$^2$ 
in Sec.~III. 

Finally, using Eqs.~(\ref{eqn:II-24}) and (\ref{eqn:II-25}), we arrive 
at the whole expressions for the imaginary part of amplitude (\ref{eqn:II-3}):
for $\gamma(L)\rightarrow \psi(L)$,
\beq
\lefteqn{{\rm Im}\ {\cal M}^L(t)=
-\frac{2\sqrt{2}\pi^2ee_cQs R}{\sqrt{N_c}}
\int_0^1d\alpha \alpha(1-\alpha)}
\nonumber\\
&\times& \alpha_s(\Qbar^2+\kvec^2)xG(x,\Qbar^2+\kvec^2)
\nonumber\\
&\times& \int_0^{\infty}\frac{d\Kvec^2}{16\pi^2}
\frac{\phi_\psi^*(\alpha,\Kvec)}{\left(\Kvec^2+\Qbar^2\right)^2}
\nonumber\\
&&\times\left(1+\frac{\Kvec^2+m_c(M_\psi+m_c)}
{\alpha(1-\alpha)M_\psi^2}\right)
\nonumber\\
&\times& \left[
H_1(\alpha,\Kvec)
+\frac{2(\alpha^2+(1-\alpha)^2)(-t)}{\Kvec^2+\Qbar^2} H_2(\alpha,\Kvec)\right]\ ;
\nonumber\\
\label{eqn:II-26}
\eeq
for $\gamma(T)\rightarrow \psi(T)$,
\begin{eqnarray}
\lefteqn{{\rm Im}\ {\cal M}^T(t)
= -\frac{\sqrt{2}\pi^2ee_cs R}{\sqrt{N_c}M_\psi^2}
\int_0^1\frac{d\alpha}{\alpha(1-\alpha)}}
\nonumber\\
&\times&
\alpha_s(\Qbar^2+\kvec^2)xG(x,\Qbar^2+\kvec^2)
\nonumber\\
&&\times \int_0^{\infty}\frac{d\Kvec^2}{16\pi^2}
\frac{\phi_\psi^*(\alpha,\Kvec)}{\left(\Kvec^2+\Qbar^2\right)^2}
\nonumber\\
&\times&
\Biggl[-m_c\left\{(M_\psi+m_c)m_c+\alpha(1-\alpha)M_\psi^2\right\} 
H_1(\alpha,\Kvec)\Biggr.
\nonumber\\
&& +\ 2\left\{(\alpha^2+(1-\alpha)^2)M_\psi+m_c\right\}
\frac{\Kvec^2\Qbar^2}{\Kvec^2+\Qbar^2}
\nonumber\\
&+&\frac{2(\alpha^2+(1-\alpha)^2)(-t)}{\Kvec^2+\Qbar^2}
\nonumber\\
&&\times \Biggl\{
-m_c\left((M_\psi+m_c)m_c+\alpha(1-\alpha)M_\psi^2\right) H_2(\alpha,\Kvec)
\Biggr.
\nonumber\\
&&\qquad+3\left((\alpha^2+(1-\alpha)^2)M_\psi+m_c\right)
\nonumber\\
&&\qquad\times
\left.\left.\frac{\Kvec^2\Qbar^2}{\Kvec^2+\Qbar^2} H_1(\alpha,\Kvec)
\right\}\right].
\label{eqn:II-27}
\end{eqnarray}
%
Here, we define 
\begin{eqnarray}
H_1(\alpha,\Kvec) &=& 1-\frac{2\Qbar^2}{\Kvec^2+\Qbar^2},
\nonumber\\
H_2(\alpha,\Kvec) &=& 1-\frac{6\Qbar^2}{\Kvec^2+\Qbar^2}+\frac{6\Qbar^4}
{\left(\Kvec^2+\Qbar^2\right)^2}.
\label{eqn:II-28}
\end{eqnarray}

To determine the nonperturbative part of the charmonium wave functions, 
$\phi_\psi(\alpha,\Kvec)$, we use the wave functions obtained by solving
Schr\"{o}dinger equation with the realistic potential \cite{FKS,SHIAH}.
We adopt the Cornell potential model 
with the corresponding quark masses $m_c=1.5$ GeV \cite{QR}, 
giving a good description of the charmonium wave functions.
Here, we rewrite the non-relativistic wave function, $\phi_\psi^{NR}$, 
originally obtained in terms of three momenta, 
as a function of the LC variables, $\phi_\psi(\alpha,\Kvec)$, 
by simple kinematical replacement \cite{FKS,SHIAH}. 
For the diagonal gluon distribution $G$ in Eqs.~(\ref{eqn:II-26}), (\ref{eqn:II-27}), 
we employ Gl\"uck-Reya-Vogt parametrization for the next-to-leading 
order fits \cite{GRV95}.

Using Eqs.~(\ref{eqn:II-26}) and (\ref{eqn:II-27}), 
let us construct $B^{dip}_V(t)$ in Eq.~(\ref{eqn:I-3}), 
which comes from the upper part
describing the transition $\gamma^{(*)}\rightarrow V$ in Fig.~ \ref{fig:fig1}. 
One should note that, in our approximation up to the first order of $t$, 
$B^{dip}(t)$ is a function of $Q^2$ and $W$, whereas independent of $t$.   
Conventionally, we denote the imaginary part of the amplitudes as
${\rm Im}{\cal M}_{L(T)}={\cal A}^{Im}_{L(T)}+(-t)\tilde{\cal A}^{Im}_{L(T)}$,
where ${\cal A}^{Im}_{L(T)}$ corresponds to the amplitude independent of $t$
in Eqs.~(\ref{eqn:II-26}) and (\ref{eqn:II-27}), and then $\tilde{\cal A}^{Im}_{L(T)}$ 
denotes the amplitude proportional to $t$. 
Similarly, for the real part, 
${\rm Re}{\cal M}_{L(T)}={\cal A}^{Re}_{L(T)}+(-t)\tilde{\cal A}^{Re}_{L(T)}$.
Using these amplitudes to the first order of $t$, 
$B^{dip}_V(W,Q^2)$ is given by
\beq
B^{dip}_V(W,Q^2) = -2\frac{\sum_{i=L,T}
\left[{\cal A}^{Im}_i\tilde{\cal A}^{Im}_i
+{\cal A}^{Re}_i\tilde{\cal A}^{Re}_i\right]}
{\sum_{i=L,T}\left[\left({\cal A}^{Im}_i\right)^2
+\left({\cal A}^{Re}_i\right)^2\right]} .
\label{eqn:II-28}
\eeq

\section{Numerical results and comparison with data}
\begin{figure}[h]
\vspace*{0cm}
\begin{center}
\hspace*{0cm}
\psfig{file=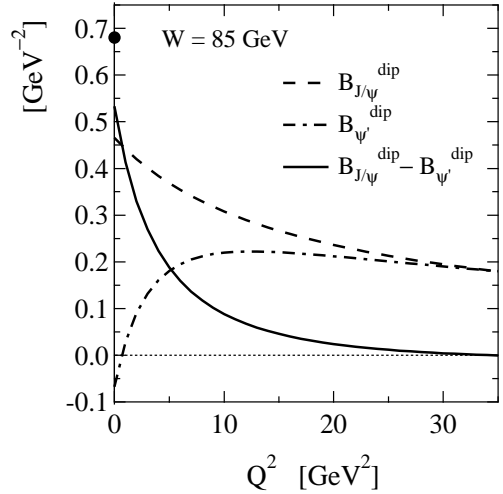,width=7.5cm,height=7cm}
\end{center}
\vspace{0cm}
\caption{The slope parameters $B^{dip}_{J/\psi}$, $B^{dip}_{\psi'}$, 
and $B^{dip}_{J/\psi}- B^{dip}_{\psi'}$, plotted as a function of $Q^2$
at the fixed energy $W=85$ GeV. 
A blob at $Q^2=0$ represents $B_{J/\psi}- B_{\psi'}$ 
from the data \cite{H102} for a reference.}
\label{fig:fig3}
\end{figure}

%
In Fig.~\ref{fig:fig3}, we show the results 
of the dipole contribution (\ref{eqn:II-28}) to the slope parameters, 
$B^{dip}_V$, as a function of $Q^2$.
Those results are plotted as dashed line for $B^{dip}_{J/\psi}$, 
dash-dotted line for $B^{dip}_{\psi'}$,  
and solid line for their difference $B^{dip}_{J/\psi}- B^{dip}_{\psi'}$,
at the fixed energy $W=85$ GeV.
%
%
$B^{dip}_{J/\psi}$ has the maximum value 
of $B^{dip}_{J/\psi}\simeq 0.48$ GeV$^{-2}$ 
and decreases with increasing $Q^2$, 
while $B^{dip}_{\psi'}$ has the negative value near $Q^2=0$, 
$B^{dip}_{J/\psi}\simeq -0.07$ GeV$^{-2}$, and 
rapidly increases with increasing $Q^2$, taking the positive value.
This feature changing the sign of the $t$-slope with $Q^2$
appears only in the $\psi'$ production. 
This is an indication of the node inherent 
in the radial direction of the $2S$ wave function. 
At large $Q^2$, however, they approach each other, 
and then their difference reaches almost zero, 
because the size of the $c\bar{c}$-dipole state 
is squeezed with increasing $Q^2$ and 
the amplitudes are evaluated near the origin 
of the charmonium wave functions.
We find the maximum value of the difference at $Q^2=0$, which is
$B^{dip}_{J/\psi}- B^{dip}_{\psi'}\simeq 0.53$ GeV$^{-2}$.
Using Eq.~(\ref{eqn:I-3a}), we can directly compare the result 
with the experimental data.  
In fact, it turns out that 
our result is consistent with H1 data \cite{H102}, which
is $B_{J/\psi}- B_{\psi'}\simeq 0.68$ GeV$^{-2}$
showed as a blob in Fig.~\ref{fig:fig3}, in the sign 
and the magnitude.

\begin{figure}[h]
\vspace*{0cm}
\begin{center}
\hspace*{0cm}
\psfig{file=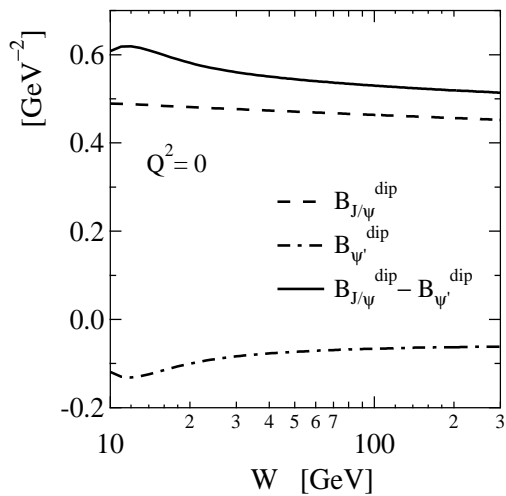,width=7.5cm,height=7cm}
\end{center}
\vspace{0cm}
\caption{
The $W$-dependence of the slope parameters 
$B^{dip}_{J/\psi}$, $B^{dip}_{\psi'}$, and $B^{dip}_{J/\psi}- B^{dip}_{\psi'}$,
for photoproductions, compared with ZEUS data \cite{ZEUS02}.
}
\label{fig:fig4}
\end{figure}

Similar behavior of the $t$-slope difference as a function of $Q^2$
was also confirmed in the analysis of \cite{NNPZZ} 
for the transverse polarization.
Their results show that the difference is 
about 0.25 GeV$^{-2}$ at $Q^2=0$ and $W=100$ GeV, 
which is about 2 times smaller than our result,
and falls down more slowly with increasing $Q^2$.
In Fig.~\ref{fig:fig4}, we plot
the slope parameters $B^{dip}_{J/\psi}$, $B^{dip}_{\psi'}$, 
and $B^{dip}_{J/\psi}- B^{dip}_{\psi'}$, as a function of $W$
at $Q^2=0$, similar to Fig.~\ref{fig:fig3}.
%
%
Apparently, both $B^{dip}_{J/\psi}$ and $B^{dip}_{\psi'}$ show an almost flat 
behavior as a function of $W$ and their absolute values even decrease 
slightly with increasing $W$. 
As a result, their difference $B^{dip}_{J/\psi}- B^{dip}_{\psi'}$ 
is also almost independent of $W$.  
Such insensitive behaviors to $W$ are found also at moderate $Q^2$.
On the other hand, the $t$-slope difference in \cite{NNPZZ} shows
stronger $W$-dependence as $W$ increasing:
their result decreases by a factor $\sim 2$ from 
the moderate fixed-target energies to the HERA collider energies.

\begin{figure}[h]
\vspace*{0cm}
\begin{center}
\hspace*{0cm}
\psfig{file=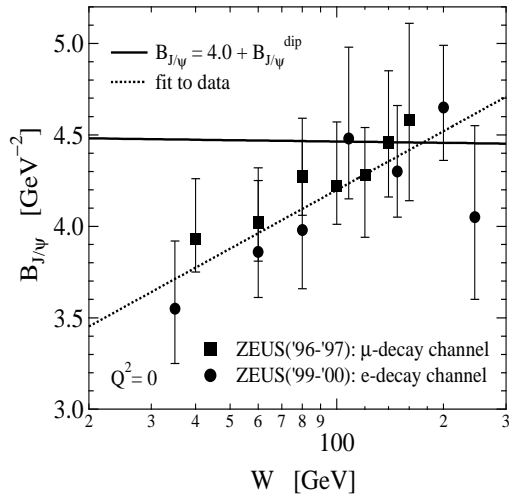,width=7.5cm,height=7cm}
\end{center}
\vspace{0cm}
\caption{The $W$-dependence of the $t$-slope $B_{J/\psi}$
for the elastic $J/\psi$ photoproduction. 
The data was measured in two leptonic decay channels, 
$J/\psi\rightarrow \mu^+\mu^-,e^+e^-$ \cite{ZEUS02}. 
The solid line shows our result
$B_{J/\psi}(t=0)=4+B^{dip}_{J/\psi}$ in Eq.~(\ref{eqn:I-3}).
The dotted line represents the fit to the data: 
$B_V(W)=B_0+4\alpha'_{\bf P}\cdot {\rm ln}\,(W/90\,{\rm GeV})$
with $B_0=4.15$ GeV$^{-2}$
and $\alpha'_{\bf P}=0.116$ GeV$^{-2}$ \cite{ZEUS02}.
}
\label{fig:fig5}
\end{figure}

We can now compare the result of $B^{dip}_{J/\psi}$ with ZEUS data \cite{ZEUS02}
by combining with the $t$-slope arising from the two-gluon form factor 
as in Eq.~(\ref{eqn:I-3}). 
Here, we simply set $t=0$ for the first term of (\ref{eqn:I-3}), and
then it adds a constant factor of $B_N$ to 
$B^{dip}_{J/\psi}$ as $B_{J/\psi}(t=0)=4.0+B^{dip}_{J/\psi}$.
Using this relation, we find that the contribution of the $c\bar{c}$ state 
to the $t$-slope is not negligible, because the size can reach about 10 $\%$ 
of the whole at $Q^2=0$ for the $J/\psi$ production. 
Similarly for the $\psi'$ production, it turns out that 
the contribution of the $c\bar{c}$ state
becomes largest (about 5 $\%$) in the vicinity of $Q^2\sim 10$ GeV$^2$.
The result of $B_{J/\psi}(W)$ is shown in Fig.~\ref{fig:fig5} with the data measured 
for the muon and electron decay channels of $J/\psi$ separately.
%
%
The data indicate the increase of the $t$-slope with $W$ 
and this energy-dependent $t$-slope fits well to the data 
as a form of $B_V(W)=B_0+4\alpha'_{\bf P}\cdot {\rm ln}\,(W/90\,{\rm GeV})$ 
with $B_0=4.15$ GeV$^{-2}$
and $\alpha'_{\bf P}=0.116$ GeV$^{-2}$ \cite{ZEUS02}.  
%
Our $t$-slope almost independent of $W$ is in agreement with the data only 
in the range of $W=100\sim 200$ GeV.
The reason for this behavior (solid line)  
is easily explained in our formulation:
$B^{dip}_V(W)$ giving only the $W$-dependence in Eq.~(\ref{eqn:I-3}) 
has the form $B^{dip}_V\propto {\cal A}\tilde{\cal A}/{\cal A}^2$ 
in a simplest description, 
where ${\cal A}$($\tilde{\cal A}$) denotes the imaginary or real parts 
of the amplitude with either longitudinal or transverse polarizations. 
Then, the amplitudes has a common factor $xG(x)$ in the integrands.
Usually, at HERA energies, the low-$x$ behavior 
of the gluon distribution $xG(x)$ is expected to be $xG(x)\sim x^{-\lambda}$ 
with $\lambda\simeq 0.2$, 
which is obtained from the analysis of the inclusive deep-inelastic scattering data. 
Here, $x\sim M_\psi^2/W^2$ at $Q^2=0$
and the $Q^2$-scale of $G$ is set to $Q^2\simeq m_c^2$. 
Following this simple parametrization for $xG(x)$, $B^{dip}_V$ is 
completely independent of $W$. 
In fact, $x$ and the scale $\Qbar^2+\kvec^2$
have $\alpha$, $\kvec$-dependences in our formulation.  
The corrections, however, give only a weak dependence of $B^{dip}_V$ 
on the energy $W$, 
illustrated in Fig.~\ref{fig:fig5}.
Thus, the energy-dependence of the $t$-slope, so called "shrinkage", 
observed by the ZEUS experiment, seems 
to need more sophisticated treatments for the  $W$ dependence. 
%
For example, we may consider two possibilities as the origin of the $W$-dependence, 
which we have not taken into account here: 
one possible effect is the Gribov diffusion, which is known as a rapid expansion 
in the transverse size of the $c\bar{c}$-dipole state, 
created from a photon fluctuation as $W$ increases. 
This phenomenon leads to the shrinkage of a diffraction peak, 
which can be interpreted as an increase of the interaction radius \cite{BFGMS};
another effect is the increase of the $c\bar{c}$-dipole scattering 
off the peripheral pion-cloud of a nucleon with increasing $W$. 
As suggested in \cite{FS}, at relatively low $W$($\sim 10$ GeV) 
such as the fixed-target energies, the contribution of the pions 
to the gluon distribution is not important. 
It is because the pions carry a smaller fraction of the nucleon momentum,
so that the gluons inside the pions have much smaller momentum fraction 
than $x\sim 0.1$, 
which is the typical value of the momentum fraction at the energies. 
In this case, the authors of \cite{FS} predict 
that the corresponding two-gluon form factor (\ref{eqn:II-a})
is close to the axial form factor of the nucleon with $m_{2g}^2\simeq 1$ GeV$^2$.
%
%
At higher collider energies of HERA (typically, $W\sim 100$ GeV), 
where it reaches $x\ll 0.01$, one should take into account 
the contribution from the pion-cloud. 
At high $W$, therefore, the form factor is expected to approach 
the electromagnetic one with $m_{2g}^2\simeq 0.71$ GeV$^2$ \cite{FS}. 
%
%

%
In order to check how the mass scale changes with $W$ in our approach, 
we try to reproduce 
the HERA data of the elastic $J/\psi$ photoproduction 
by changing a value of the mass scale.  
We neglect a contribution from the Gribov diffusion, 
because the pQCD analysis is inapplicable for a large size configuration 
due to the diffusion. 
%
%

\begin{figure}[h]
\vspace*{0cm}
\begin{center}
\hspace*{0cm}
\psfig{file=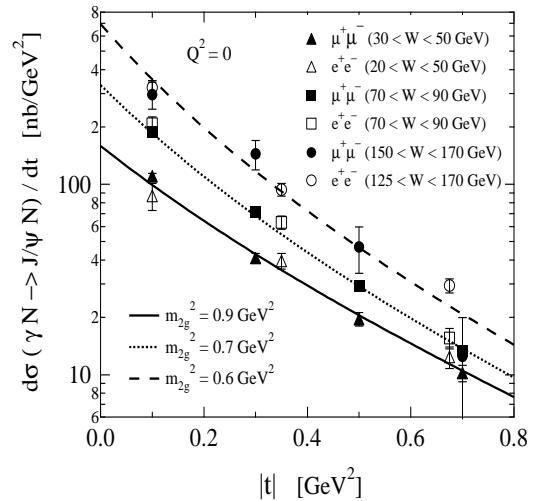,width=7.5cm,height=7cm}
\end{center}
\vspace{0cm}
\caption{The differential cross section of elastic
$J/\psi$ photoproduction, $\gamma+N\rightarrow J/\psi+N$,
as a function of $|t|$, compared with ZEUS data \cite{ZEUS02} 
measured in two leptonic decay channels, 
$J/\psi\rightarrow \mu^+\mu^-, e^+e^-$.
The data are plotted for three representative ranges of $W$ 
for respective decay channels, as written in the figure. 
The solid, dotted and dashed curve show our results calculated 
in three corresponding $W$ ranges, $30<W<50$ GeV, $70<W<90$ GeV and $150<W<170$ GeV.
Then good fits to the data require that the squared mass scale has
the values 0.9, 0.7, 0.6 GeV$^2$ in respective $W$ regions.
}
\label{fig:fig2}
\end{figure}

Resulting $t$-dependence of the $J/\psi$ cross section is shown in Fig.~\ref{fig:fig2}.  
%
It is compared with ZEUS data \cite{ZEUS02} for
three representative ranges of $W$, {\it i.e.},
$30<W<50$ GeV, $70<W<90$ GeV and $150<W<170$ GeV for the $\mu^+\mu^-$ decay of $J/\psi$,
and $20<W<50$ GeV, $70<W<90$ GeV and $125<W<170$ GeV for the $e^+e^-$ decay.
They are calculated over the kinematic range $|t| < 0.8$ GeV$^{-2}$,
in which our formulation could be applicable.
%
%
Here, we set a parameter $R=1.2$ for the skewedness effect of 
the off-diagonal gluon distribution, 
which changes only the normalization of the cross section 
and thus does not affect the dependence of cross section on $t$.
Further, to increase the overall magnitude of the cross section, 
we carry out a rescaling of $Q^2$ in the gluon distribution
and the strong coupling constant \cite{rescale}, 
following the work of \cite{FKS}.
%
%
%
It is also stressed that our naive treatment for the rescaling
is almost insensitive to the $t$-slope, but it enhances
only the normalization of the cross section about $30\sim 40$ $\%$.

The solid, dotted, and dashed curves represent our calculations
corresponding to each $W$ range, $30<W<50$ GeV, $70<W<90$ GeV and $150<W<170$ GeV. 
They show an exponential decrease very close to the data 
with increasing $|t|$, 
when we take $m_{2g}^2=0.9, 0.7, 0.6$ GeV$^2$ in respective $W$ bins.
Hence, to obtain a good agreement with the data, 
we find that the mass scale should decrease 
with increasing $W$ at HERA energies.  
This is qualitatively consistent with the suggestion of \cite{FS}
mentioned above.

\section{Summary and discussion}
%
To summarize, in the LLA of pQCD we have formulated $t$-dependences 
of the differential cross section for the diffractive ("elastic") 
charmonia ($J/\psi$ and $\psi'$) 
photo- and electroproductions at low $|t|$($<1$ GeV$^{-2}$). 
To deal with the production of the radially excited $2S$ state ($\psi'$) 
together with that of $J/\psi$, we have employed the charmonium LCWF, 
properly including the sub-leading effects due to Fermi motion, 
based on a technique developed in \cite{HST2,HST3}.   
%
Following QCD factorization, the two-gluon form factor of the nucleon 
is assumed to be process-independent and we define it as a function 
of only $t$ scaled by the relevant squared mass in a dipole form.
By assuming an exponential form of the differential cross section, 
following the standard experimental fit so far, 
we have calculated the diffractive $t$-slope, $B_V(t)$, 
over the $|t|$ range of less than 1 GeV$^2$.

The dipole part, $B^{dip}_V$, shows that the $Q^2$ 
dependence presents a distinct difference  between $J/\psi$ and $\psi'$, 
reflecting the properties of wave functions, 
{\it i.e.}, a node effect for the $2S$ state.
The former monotonically decreases with $Q^2$, 
while the latter has a negative value at small $Q^2$ 
and rapidly increases with $Q^2$, keeping $B^{dip}_{J/\psi} > B^{dip}_{\psi'}$
in the realistic $Q^2$ region of HERA.
Then, the difference $B_{J/\psi}-B_{\psi'}$ shows a strong $Q^2$ dependence 
that tends to almost zero with increasing $Q^2$ 
associated with squeezing of the dipole,
after achieving a maximum of $0.53$ GeV$^{-2}$ at $Q^2=0$ and $W=85$ GeV.
This value at $Q^2=0$ is consistent with HERA data.
It is noted that, to a good approximation,
the difference should be free from a contribution 
of the Gribov diffusion, 
because such a contribution should be incorporated 
into the initial photon part in QCD factorization 
and it is expected to contribute at the same level for $J/\psi$ and $\psi'$. 

The $W$-dependences of $B^{dip}_V$ indicate almost a flat behavior at $Q^2=0$
(or also moderate $Q^2$), contrary to a shrinkage of $t$-slope 
observed at ZEUS.
At small $x$, mainly two effects are expected to be the origin 
for such $W$-dependences of the $t$-slope: 
one comes from the dipole part, $\it i.e.$, the Gribov diffusion;
another is due to the target nucleon, $\it i.e.$, 
the effect of hard probe scattering off the peripheral pion-cloud of a nucleon.
We incorporated the latter phenomenon through the change of the mass scale 
appearing in the nucleon form factor, by fitting 
to ZEUS data for differential cross section of $J/\psi$ photoproductions 
at several $W$ ranges, although we neglected the former phenomenon for simplicity.
To get a good fit to the data, we find significant decrease of the mass scale 
with increasing $W$, {\it e.g.}, 
$0.6 < m_{2g}^2 < 0.9$ GeV$^2$ in the region of $20 < W < 170$ GeV.
This mass scale decreasing with $W$ will provide us 
with information concerning gluon distributions of the nucleon 
through a probe of $c\bar{c}$-dipole.
Further detailed study on it might be useful 
for investigating not only the gluon distribution of the pion-cloud 
on the surface of nucleon, but also the non-linear property of gluons inside the nucleon, 
which will become more important with higher $W$.

On the other hand, the difference of $t$-slopes 
shows a behavior almost independent of $W$. 
This quantity is most likely free from the Gribov diffusion,
as mentioned above. 
%
%
Unfortunately, the $W$-dependence of $B_{\psi'}$ has been not yet observed so far.
However, its observation has some importance
and we would request early realization of such an observation.
If the experiments observe somewhat the $W$ dependence 
for the difference of $t$-slopes, contrary to our prediction, 
then we might guess the following two origins:
\begin{enumerate}
\item 
The Gribov diffusion might introduce a large size effect upon the $c\bar{c}$-dipole
at high $W$, thereby influencing the final hadronization dependence. 
It will give important information about the mechanism of the diffusion.
\item
Otherwise, we could doubt the hypothesis for a universal form 
of nucleon form factor based on the factorization, which is very challenging. 
\end{enumerate}

Finally, we stress that the finite size effect of the $c\bar{c}$-dipole 
is not negligible for the $t$-slope of charmonia electroproductions 
at the HERA energies and $Q^2 < 30$ GeV$^2$.
Combining both data from $J/\psi$ and $\psi'$ productions,
we can investigate the pure contribution of the dipole part in detail.
Such a precise study will finally make it important 
to extract the detailed information of the gluon distribution in the nucleon 
by making use of the diffractive charmonium photoproductions.
Recently, a fascinating study on the energy-dependent $t$-slope 
was reported in \cite{KT}, using the dipole saturation model 
with an impact parameter dependence. A part of the $W$ dependence 
could also arise from the non-linear dynamics of the saturation 
for the gluon distribution in the target nucleon. 
The analysis including such an effect we consider to be interesting 
and is planned for one of our future works.
%

\acknowledgements
We are grateful to K.~Tanaka for a lot of useful discussions 
during the initial stage of this work
and T.~Burch for the careful reading of the manuscript.
A.H. is supported by Alexander von Humboldt Research Fellowship.



\begin{references}
\bibitem{rescale}In numerical calculation, this rescaling is simply done by adopting 
$Q^2_{eff}=2(\alpha(1-\alpha)Q^2+m_c^2)$, which is different 
from relations used in \cite{FKS}, but it gives results close to \cite{FKS}
quantitatively \cite{HST3}.
\end{references}
\end{document}